\documentclass[11pt,a4paper]{article}

\usepackage{epsfig,amsmath,amssymb}

\def\be{\begin{equation}}
\def\ee{\end{equation}}
\def\lp{\left(}
\def\rp{\right)}
\def\lb{\left[}
\def\rb{\right]}

\begin{document}

\title{Thin-shell wormholes in dilaton gravity} 
\author{Ernesto F. Eiroa$^{1,}$\thanks{e-mail: eiroa@iafe.uba.ar}, 
Claudio Simeone$^{2,}$\thanks{e-mail: csimeone@df.uba.ar}\\
{\small $^1$ Instituto de Astronom\'{\i}a y F\'{\i}sica del Espacio, C.C. 67, 
Suc. 28, 1428, Buenos Aires, Argentina}\\
{\small $^2$ Departamento de F\'{\i}sica, Facultad de Ciencias Exactas y 
Naturales,} \\ 
{\small Universidad de Buenos Aires, Ciudad Universitaria Pab. I, 1428, 
Buenos Aires, Argentina}} 

\maketitle

\begin{abstract}
In this work we construct charged thin-shell Lorentzian wormholes in dilaton 
gravity. The \textit{exotic matter} required for the construction is localized 
in the shell and the energy conditions are satisfied outside the shell. The 
total amount of \textit{exotic matter} is calculated and its dependence with 
the parameters of the model is analysed.\\ 

\noindent 
PACS number(s): 04.20.Gz, 04.50.+h, 04.40.Nr

\end{abstract}

\section{Introduction}\label{intro}

Since the leading paper by Morris and Thorne \cite{motho} the study of 
traversable Lorentzian wormholes has received considerable attention. These 
objects are solutions of the equations of gravitation that have two regions 
(of the same universe or may be of two separate universes 
\cite{motho, visser}) connected by a throat. For static wormholes, the throat 
is defined as a two-dimensional hypersurface of minimal area that must satisfy 
a \textit{flare-out} condition \cite{ hovis1}.  All traversable wormholes 
include  \textit{exotic matter}, which violates the null energy condition 
(NEC) \cite{motho, visser, hovis1, hovis2}. Interesting discussions about the 
energy conditions and wormholes are given in the essays by Barcel\'{o} and 
Visser \cite{bavis} and by Roman \cite{roman}. Recently, there has been a 
growing interest in quantifying the amount of \textit{exotic matter} present 
around the throat. Visser \textit{et al.} \cite{viskardad} showed that the 
\textit{exotic matter} can be made infinitesimally small by appropriately 
choosing  the geometry of the wormhole, and Nandi \textit{et al.} 
\cite{nandi1} proposed a precise volume integral quantifier for the average 
null energy condition (ANEC) violating matter. This quantifier was 
subsequently used by Nandi and Zhang \cite{nandi2} as a test of the physical 
viability of traversable wormholes. A well studied class of wormholes are 
thin-shell ones, which are constructed by cutting and pasting two manifolds 
\cite{visser, mvis} to form a geodesically complete new one with a throat 
placed in the joining shell. Thus, the \textit{exotic matter} needed to build 
the wormhole is located at the shell and the junction-condition formalism is 
used for its study. Poisson and Visser \cite{poisson} made a linearized 
stability analysis under spherically symmetric perturbations of a thin-shell 
wormhole constructed by joining two Schwarzschild geometries. Later, Eiroa 
and Romero \cite{eirom} extended the linearized stability analysis to 
Reissner--Nordstr\"{o}m thin-shell geometries, and Lobo and Crawford 
\cite{lobo-craw} to wormholes with a cosmological constant. Lobo, with the 
intention of minimizing the \textit{exotic matter} used, matched a static and 
spherically symmetric wormhole solution to an exterior vacuum solution with a 
cosmological constant, and he calculated the surface stresses of the resulting 
shell \cite{lobo1}, and the total amount of \textit{exotic matter} using a 
volume integral quantifier \cite{lobo2}. Cylindrically symmetric thin-shell 
wormhole geometries associated to gauge cosmic strings have also been treated 
by the authors of the present work \cite{eisi}. \\

In this article we study spherical thin-shell wormholes in dilaton gravity, 
that is, wormholes constructed by cutting and pasting two metrics 
corresponding to a charged black hole which is a solution of low energy string 
theory, with dilaton and   Maxwell fields, but vanishing antisymmetric field 
and all other gauge fields set to zero \cite{GHS,gib,mae}. Lorentzian four 
dimensional traversable wormholes in the context of low energy effective 
string theory or in Einstein gravity with a massless scalar field were already 
proposed by Kim and Kim \cite{kim-kim}, Vollick \cite{vol}, Barcel\'{o} and 
Visser \cite{BV}, Bronnikov and Grinyok \cite{BG}, Armend\'{a}riz-Pic\'{o}n 
\cite{armen}, Graf \cite{graf}, Nandi and Zhang \cite{NZ}, and Nandi 
\textit{et al.} \cite{NZK}, but, differing from our construction, in these 
works the energy conditions were violated in a non compact region. Moreover, 
as we consider a non vanishing charge, this allows for a comparison with the 
thin-shell wormhole associated to the  Reissner--Nordstr\"{o}m  geometry 
\cite{eirom}. We focus on the geometry of these objects and we do not intend 
to give any explanation about the mechanisms that might supply the 
\textit{exotic matter} to them; however, we shall analyse in detail the 
dependence of the total amount of \textit{exotic matter} with the parameters 
of the model. 
Thin-shell charged dilaton wormholes are constructed in Sec. \ref{tswh}, the 
energy conditions are studied in Sec. \ref{ec}, and the results obtained are 
summarized in Sec. \ref{conclu}. Throughout the paper we use units such as 
$c=G=1$.\\

\section{Thin-shell wormholes in low energy string  gravity}\label{tswh}
 
Following  Ref. \cite{GHS}, here we shall consider low energy string theory 
with Maxwell field, but with  all other gauge fields and antisymmetric field 
set to zero \cite{polch}. Then   the effective theory includes three fields: 
the spacetime metric $g_{\mu\nu}$, the (scalar) dilaton field $\phi$, and the 
Maxwell field $F^{\mu\nu}$. The corresponding action  in the Einstein frame 
\cite{gas} reads
\be S=\int d^4x\sqrt{-g}\left(-R+2(\nabla \phi)^2+e^{-2b\phi}F^2\right),
\label{emd}
\ee
where $R$ is the Ricci scalar of spacetime. We have also included a parameter 
$b$ in the coupling between the dilaton and the Maxwell field, as is done in 
Ref. \cite{GHS} to allow for a more general analysis; this parameter will be 
restricted within the range $0\leq b\leq 1$. The interpretation of the 
gravitational aspects of the theory is clear in the Einstein frame: In 
particular, the condition $\delta S=0$ imposed  on the action (\ref{emd}) 
leads to the Einstein equations with the dilaton and the Maxwell field as 
the source. Within the context of our work, this is of particular interest, 
as it allows for a comparison with the results already known for 
Einstein--Maxwell theory. The field equations yielding from the action 
(\ref{emd}) are
\be
\nabla _{\mu }\lp e^{-2b\phi }F^{\mu \nu }\rp =0,
\ee
\be
\nabla ^{2}\phi +\frac{b}{2}e^{-2b\phi }F^{2}=0,
\ee
\be
R_{\mu \nu }=2\nabla _{\mu }\phi \nabla _{\nu }\phi +2e^{-2b\phi }
\lp F_{\mu \alpha }F_{\nu }^{\; \alpha }-\frac{1}{4}g_{\mu \nu }F^{2}\rp . 
\label{feq}
\ee
These equations admit black hole spherically symmetric solutions, with metric 
in curvature (Schwarzschild) coordinates \cite{GHS,mae}:
\be 
ds^2=-f(r)dt^2+f^{-1}(r)dr^2+h(r)(d\theta ^2+\sin^2\theta d\varphi^2), 
\label{metric}
\ee
where
\begin{eqnarray}f(r) & = &\lp 1-\frac{A}{r}\rp\lp 1-\frac{B}{r}\rp^{\frac{1-b^2
}{1+b^2}}, \nonumber\\
h(r) & = &r^2\lp 1-\frac{B}{r}\rp^{\frac{2b^2}{1+b^2}}.\label{fyh}
\end{eqnarray}
The constants $A,B$ and the parameter $b$ are related with the mass and charge 
of the black hole by
\begin{eqnarray} M & =&\frac{A}{2}+\lp{\frac{1-b^2}{1+b^2}}\rp\frac{B}{2}, 
\nonumber\\
Q&=&\sqrt{\frac{AB}{1+b^2}}.\label{mq}
\end{eqnarray}
In the case of electric charge, the electromagnetic field tensor has non-null 
components $F_{tr}=-F_{rt}=Q/r^{2}$, and the dilaton field is given by 
$e^{2\phi}=\lp 1-B/ r\rp^{2b/(1+b^2)}$,
where the asymptotic value of the dilaton $\phi_{0}$ was taken as zero. For 
magnetic charge, the metric is the same, with the electromagnetic field 
$F_{\theta \varphi}=-F_{\varphi \theta}=Q\sin \theta $ and the dilaton field 
obtained replacing $\phi $ by $-\phi $. When $b=0$, which corresponds to a 
uniform dilaton, the metric reduces to the  Reissner--Nordstr\"{o}m  geometry, 
while for $b=1$, one obtains $f(r) = 1-2M/r$, $h(r) = r^2\lb 1-Q^2/ (Mr)\rb$.
In what follows, we shall consider the generic form of the metric (\ref{fyh}), 
with $0\le b\le 1$. $B$ and $A$ are, respectively, the inner and outer 
horizons of the black hole; while the outer horizon is a regular event horizon 
for any value of $b$, the inner one is singular for any $b\neq 0$.\\ 

From the geometry given by Eqs. (\ref{metric}) and (\ref{fyh}) we  take two 
copies of the region with $r\geq a$: $\mathcal{M}^{\pm }=\{x/r\geq a\}$, 
and paste them at the hypersurface $\Sigma \equiv \Sigma ^{\pm }=
\{x/F(r)=r-a=0\}$. We consider $a>A>B$ to avoid singularities and horizons. 
Because $h(r)$ is an increasing function of $r$, the \textit{flare-out} 
condition is satisfied. The resulting construction creates a geodesically 
complete manifold $\mathcal{M}=\mathcal{M}^{+}\cup \mathcal{M}^{-}$ with two 
asymptotically flat regions connected by the throat. On this manifold we can 
define a new radial coordinate $l=\pm \int _{a}^{r}g_{rr}dr$ representing the 
proper radial distance to the throat, which is located at $l=0$;  the plus and 
minus signs correspond, respectively, to $\mathcal{M}^{+}$ and 
$\mathcal{M}^{-}$. The cut and paste procedure follows the standard 
Darmois-Israel formalism \cite{daris,mus}. Working in the orthonormal basis 
$\{e_{\hat{t}},e_{\hat{r}},e_{\hat{\theta}},e_{\hat{\varphi}}\}$ 
($e_{\hat{t}}=[f(r)]^{-1/2}e_{t }$, $e_{\hat{r}}=[f(r)]^{1/2}e_{r}$, 
$e_{\hat{\theta}}=[h(r)]^{-1/2}e_{\theta }$, 
$e_{\hat{\varphi}}=[h(r)\sin^{2} \theta ]^{-1/2}e_{\varphi }$), we have that 
the second fundamental forms (extrinsic curvature) associated with the two 
sides of the shell are:
\be
K_{\hat{\theta}\hat{\theta}}^{\pm }=K_{\hat{\varphi}\hat{\varphi}}^{\pm
}=\pm \frac{h'(a)}{2 h(a)}\sqrt{f(a)},
\label{e8}
\ee
and
\be
K_{\hat{t}\hat{t}}^{\pm }=\mp \frac{f'(a)}{2\sqrt{f(a)}} ,
\label{e9}
\ee
where a prime stands for a derivative with respect to $r$. Defining 
$[K_{_{\hat{\imath}\hat{\jmath}}}]\equiv K_{_{\hat{\imath}\hat{\jmath}
}}^{+}-K_{_{\hat{\imath}\hat{\jmath}}}^{-}$, $K=tr[K_{\hat{\imath}\hat{
\jmath }}]$ and introducing the surface stress-energy tensor 
$S_{_{\hat{\imath}\hat{\jmath} }}={\rm diag}(\sigma ,p_{\hat{\theta}},
p_{\hat{\varphi}})$, the Einstein equations on the shell (Lanczos equations)
have the form: $-[K_{\hat{\imath}\hat{\jmath}}]+Kg_{\hat{\imath}\hat{\jmath}}=
8\pi S_{\hat{\imath}\hat{\jmath}}$,
which in our case give an energy density $\sigma$ and a transverse 
pressure $p_{t}=p_{\hat{\theta}}=p_{\hat{\varphi}}$:
\be
\sigma=-\frac{1}{4\pi}\frac{h'(a)}{h(a)}\sqrt{f(a)},
\label{e11}
\ee
\be
p_{t}=\frac{\sqrt{f(a)}}{8\pi}\lb \frac{f'(a)}{f(a)}+\frac{h'(a)}{h(a)}\rb .
\label{e12}
\ee
Using the explicit form (\ref{fyh}) of the metric, we obtain
\be
\sigma=-\frac{1}{2\pi a^2}\lp 1-\frac{A}{a}\rp^{\frac{1}{2}}\lp 1-\frac{B}{a}
\rp^{\frac{1-b^2}{2+2b^2}}\lb a+\frac{b^2B}{1+b^2}\lp 1-\frac{B}{a}\rp^{-1}\rb,
\label{e13}
\ee
\be
p_{t}=\frac{1}{8\pi a^{2}}\lp 1-\frac{A}{a}\rp ^{\frac{1}{2}}\lp 1-\frac{B}{a}
\rp ^{\frac{1-b^2}{2+2b^2}}\lb 2a+A\lp 1-\frac{A}{a}\rp ^{-1}+B\lp 1-
\frac{B}{a}\rp ^{-1}\rb. \label{e14}
\ee
For $b=0$ we recover the energy density and pressure corresponding to the 
thin-shell wormhole associated with the Reissner--Nordstr\"{o}m geometry, 
obtained in Ref. \cite{eirom} (static case). 
A distinctive feature of wormhole geometries is the violation of the energy 
conditions, which is studied in detail in the next section.

\section{Energy conditions}\label{ec}

The weak energy condition (WEC) states that $T_{\mu \nu}U^{\mu }U^{\nu }\ge 0$ 
for all timelike vectors $U^{\mu }$, and in an orthonormal basis it can be put 
in the form $\rho \ge 0$ and $\rho +p_{j}\ge 0$ $\forall j$, with $\rho $ the 
energy density and $p_{j}$ the principal pressures. If the WEC is satisfied, 
the local energy density will be positive for any timelike observer. The WEC 
implies by continuity the null energy condition (NEC), defined by 
$T_{\mu \nu}k^{\mu }k^{\nu }\ge 0$ for any null vector $k^{\mu }$, which in an 
orthonormal frame takes the form $\rho +p_{j}\ge 0$ $\forall j$. In the 
general case of the wormhole constructed above we have from Eqs. (\ref{e13}) 
and (\ref{e14}), as expected,  $\sigma <0$ and $\sigma +p_{r}<0$, that is a 
shell of \textit{exotic matter} which violates WEC and NEC. The sign of 
$\sigma +p_{t}$ depends on the values of the parameters.\\

Outside the shell, starting from the field equations (\ref{feq}) in the form 
$R_{\mu \nu }-\frac{1}{2}Rg_{\mu \nu }=8\pi T_{\mu \nu }$, the stress-energy 
tensor can be expressed as a sum of two terms: 
$T_{\mu \nu }=T_{\mu \nu }^{dil}+T_{\mu \nu }^{EM}$, 
where $T_{\mu \nu }^{dil}=[ \nabla_{\mu }\phi \nabla_{\nu }\phi-
\frac{1}{2}g_{\mu \nu }( \nabla \phi) ^{2}]/(4\pi )$ is the part corresponding 
to the dilaton field, and $T_{\mu \nu }^{EM}=e^{-2b\phi }\lp 
F_{\mu \alpha}F_{\nu }^{\; \alpha}-\frac{1}{4}g_{\mu \nu }F^{2}\rp/(4\pi )$ to 
the electromagnetic field. Then, using the orthonormal basis defined above, 
the energy density $\rho=T_{\hat{t}\hat{t}}$, the radial pressure 
$p_{r}=T_{\hat{r}\hat{r}}$ and the transverse pressure 
$p_{t}=p_{\hat{\theta }}=p_{\hat{\varphi }}=T_{\hat{\theta }\hat{\theta }}=
T_{\hat{\varphi }\hat{\varphi }}$, for the dilaton field are given by
\be
\rho^{dil}=p^{dil}_{r}=-p^{dil}_{t}=\frac{b^2B^2}{8\pi (1+b^2)^2r^4}
\lp 1-\frac{A}{r}\rp \lp 1-\frac{B}{r}\rp^{-\frac{1+3b^2}{1+b^2}},
\ee
and for the electromagnetic field
\be
\rho^{EM}=p^{EM}_{t}=-p^{EM}_{r}=\frac{AB}{8\pi (1+b^{2})r^{4}}
\lp 1-\frac{B}{r}\rp ^{\frac{-2b^{2}}{1+b^{2}}}.
\ee
Because our construction yields a geodesically complete space where $r>a$ and 
we have imposed  $a>A>B$ (so that there are  no horizons), we have 
$\rho^{dil}>0$, $\rho^{dil}+p^{dil}_{r}>0$, $\rho^{dil}+p^{dil}_{t}=0$, 
$\rho^{EM}>0$, $\rho^{EM}+p^{EM}_{t}>0$ and $\rho^{EM}+p^{EM}_{r}=0$, so the 
NEC and WEC are satisfied outside the shell for both the dilaton and the 
electromagnetic fields, and therefore the \textit{exotic matter} is localized 
only in the shell.\\  

\begin{figure}[t!]
\begin{center}
\vspace{-0.5cm} 
\includegraphics[width=6cm]{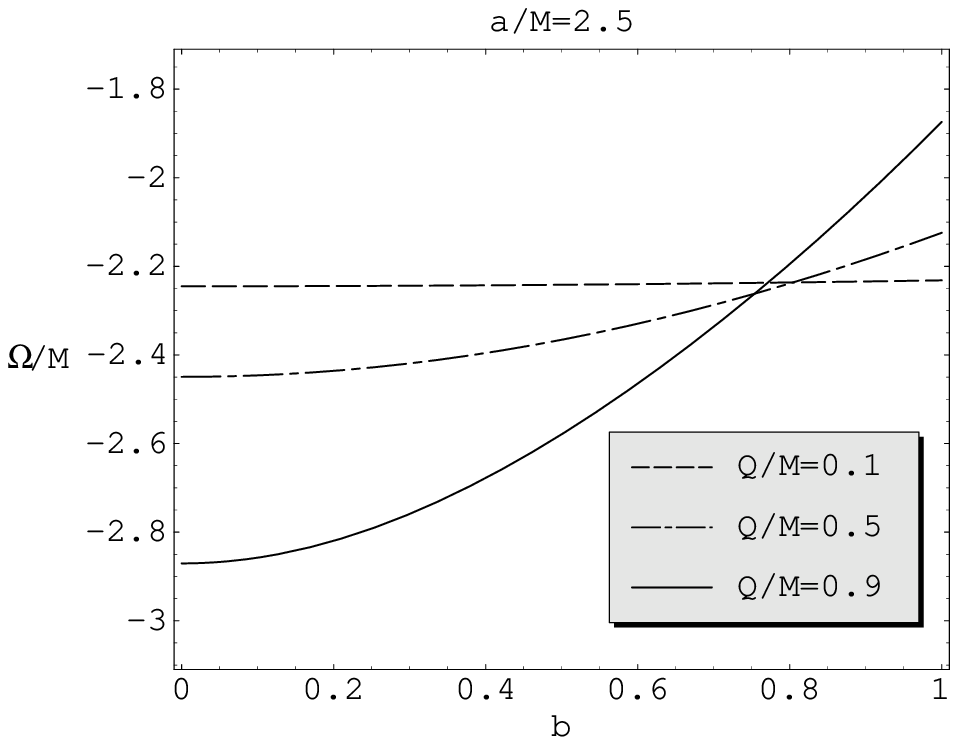}
\hspace{0.5cm}
\includegraphics[width=6cm]{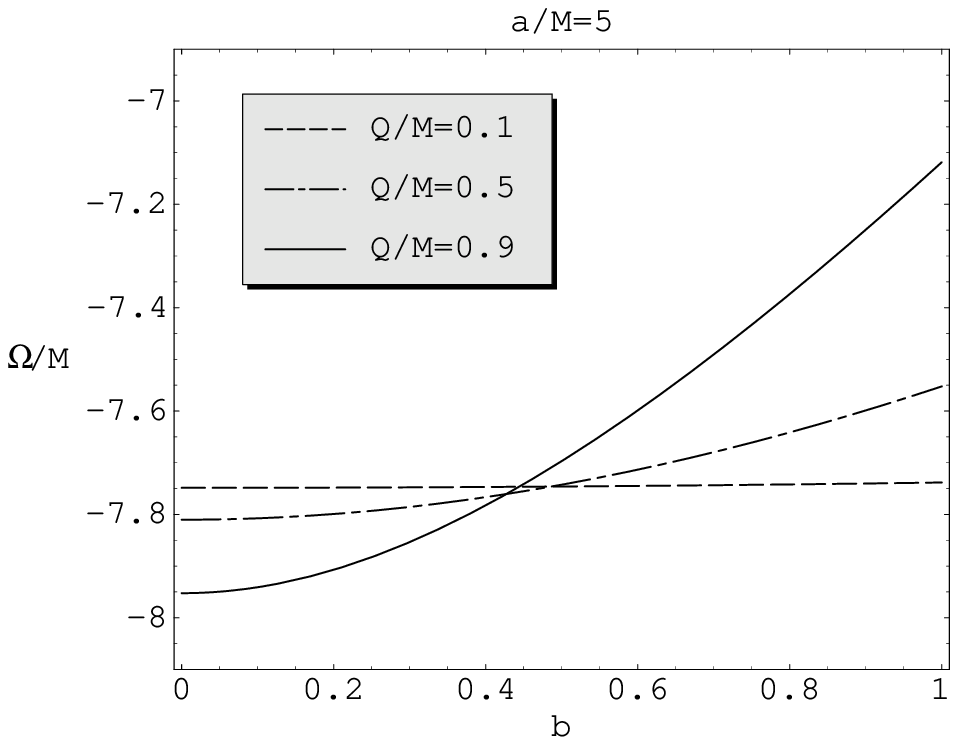}
\end{center} 
\vspace{-0.5cm} 
\caption{Total amount of \textit{exotic matter} on the shell as a function of 
the dilaton coupling parameter $b$. The curves corresponding to different 
values of charge are shown for two throat radii.}
\label{f1}
\end{figure}

The total amount of \textit{exotic matter} present can be quantified, 
following Ref. \cite{nandi1}, by the integrals $\int \rho \sqrt{-g}\, d^{3}x$, 
$\int (\rho+p_{i}) \sqrt{-g}\, d^{3}x$, where $g$ is the determinant of the 
metric tensor. The usual choice as the most relevant quantifier is 
$\Omega =\int (\rho+p_{r}) \sqrt{-g}\, d^{3}x$; in our case, introducing in 
${\cal M}$ a new radial coordinate ${\cal R}=\pm(r-a)$ ($\pm $ 
for ${\cal M}^{\pm}$, respectively) 
\be
\Omega=\int\limits_{0}^{2\pi}\int\limits_{0}^{\pi}
\int\limits_{-\infty}^{\infty}(\rho+p_{r}) \sqrt{-g}\, 
d{\cal R}d\theta d\varphi.
\ee
Taking into account that the \textit{exotic matter} only exerts transverse 
pressure and it is placed in the shell, so that 
$\rho =\delta({\cal R})\sigma $ ($\delta $ is the Dirac delta), 
$\Omega $ is given by an integral of the energy density $\sigma $ over the 
shell: 
\be
\Omega = \int\limits_{0}^{2\pi}\int\limits_{0}^{\pi} \sigma \left.
\sqrt{-g}\right|_{r=a}\,d\theta\, d\varphi=4\pi h(a)\sigma ,
\ee
 and using Eq. (\ref{e13}), we have
\be
\Omega =-2\lb a\lp 1-\frac{B}{a}\rp^{\frac{1+3b^2}{2+2b^2}}+\frac{b^2B}{1+b^2}
\lp 1-\frac{B}{a}\rp^{\frac{-1+b^2}{2+2b^2}}\rb\lp 1-\frac{A}{a}\rp
^{\frac{1}{2}}.
\label{tem}
\ee
In Fig. \ref{f1}, $\Omega  /M$ is plotted as a function of the parameter $b$ 
for different values of the charge and throat radius. We observe that: (i) For 
stronger coupling between the dilaton and the Maxwell fields (greater $b$), 
less \textit{exotic matter} is required for any given values of the radius of 
the throat, mass and charge. (ii) For large values of the parameter $b$  
(closer to $1$), the amount of \textit{exotic matter} for given radius and 
mass is reduced by increasing the charge, whereas the behaviour of $\Omega /M$ 
is the opposite for small values of $b$. It can be noticed from 
Eq. (\ref{tem}), that for $a\gg A$ there is an approximately linear dependence 
of $\Omega $ with the radius $a$ ($\Omega \approx -2 a$), and when $a$ takes 
values close to $A$, $\Omega $ can be approximated by
\be
\Omega \approx \frac{-2}{\sqrt{A}} \lb A\lp 1-\frac{B}{A}
\rp^{\frac{1+3b^2}{2+2b^2}}+\frac{b^2B}{1+b^2}\lp 1-\frac{B}{A}
\rp^{\frac{-1+b^2}{2+2b^2}}\rb \sqrt{a-A}.
\ee
The last equation shows a way to minimize the total amount of \textit{exotic 
matter} of the wormhole. For a given value of the coupling parameter $b$ and 
arbitrary radius of the throat $a$, the exotic matter needed can be reduced 
by taking $A$ close to $a$. The values of mass and charge required are then 
determined by Eq. (\ref{mq}). For $A$ close to $a$ we have $\sigma +p_{t}>0$, 
with a very small energy density and a high value of the transverse pressure. 
A consequence of reducing the amount of \textit{exotic matter} is having a 
high pressure at the throat.

\section{Conclusions}\label{conclu}

In this work, we have constructed a charged thin-shell wormhole in low energy 
string gravity. We found the energy density and pressure on the shell, and in 
the case of null dilaton coupling parameter, recovered the results obtained in 
Ref. \cite{eirom}. The \textit{exotic matter} is localized, instead of what 
happened in previous papers on dilatonic wormholes, where the energy 
conditions were violated in a non compact region. Accordingly to the proposal 
of Nandi and Zhang \cite{nandi2}, that the viability of traversable wormholes 
should be linked to the amount of \textit{exotic matter} needed for their 
construction, here we have analysed the dependence of such amount with the 
parameters of the model. We found that less \textit{exotic matter} is needed 
when the coupling between the dilaton and Maxwell fields is stronger, for 
given values of radius, mass and charge. Besides, we obtained that the amount 
of \textit{exotic matter} is reduced, when the dilaton-Maxwell coupling 
parameter is close to unity, by increasing the charge; while for low values of 
the coupling parameter, this behaviour is inverted. We have also shown how the 
total amount of \textit{exotic matter} can be reduced for given values of the 
dilaton-Maxwell coupling and the wormhole radius, by means of a suitable 
choice of the parameters.

\section*{Acknowledgments}

This work has been supported by Universidad de Buenos Aires (UBACYT X-103,
E.F.E.) and Universidad de Buenos Aires and CONICET (C.S.). Some calculations 
in this paper were done with the help of the package GRTensorII {\cite{grt}}.

\end{document}